# Comparison of Document Management Systems by Meta Modeling and Workforce Centric Tuning Measures

Harika Asılı[1] and Ömer Özgür Tanrıöver[2]

[1]Department of Computer Engineering, Ankara University, Gölbaşı Campus, Ankara

harikaasili@gmail.com

[2]Department of Computer Engineering, Ankara University, Gölbaşı Campus, Ankara

tanriover@ankara.edu.tr

## ABSTRACT

*Document management software are used widely for office paper management and related workflows. However, they have some differences from a flexibility perspective for the needs of workers. Moreover, they also differentiate in their underlying conceptual design. In this paper, a set (eight) of widely used document management systems are chosen from internet reviews [5] [6] and their conceptual models are analysed with meta modeling and flexibility tuning measures. The main objective is to compare these systems by analyzing their human-orientation and flexibility. This analysis provides useful information for organizations especially looking for document management systems which are more workforce centric.*

## 1. INTRODUCTION

Documents which are in printed format such as books, papers, forms, contracts and any related materials have been used for many years. With the evolution of information and computer systems, these documents have been managed by computer-based document management systems. A document management system (DMS) can be defined as a computer system that is used to store, track, and retrieve electronic documents. [2] There are various office types for different departments at business world. Therefore, there are many document management products in the market. These document management systems can be used for both individual paper management and business process management. On the other hand, document management systems which are used for business process management have workflow management tools. In fact, the document management systems of today are advanced workflow management system. A workflow management system (WFMS) is a software product which provides specification, execution and control of business processes [17]. All of documents can be as aligned to workflows in order to control business process in document management system.

The main objective in comparing these systems, by using meta models and analyzing their human-orientation and flexibility is based on tuning measures [4]. Firstly, document management systems are investigated from internet review websites [5] [6]. Selected document management systems are downloaded, configured and installed. These are M-Files [7] , Blue-Doc Point[8] , Speedy Organizer[9] , Iso Tracker[10] , Paperport[11] , File Center Pro[12] , Doc-Point[13] and Paperless[14] document management systems.

Some of these systems are used individually and others are used for business process management. Flexibility measures for document management systems are adapted from "tuning measures" proposed by Vanderfeesten [4]. There were twenty one tuning measures for human orientation and flexibility. These software systems are examined according to these.

## 2. DMS & WFMS EVOLUATION BASED ON META MODELLING APPROACH

Eight different document management software products are analyised. These software systems have meta model structures which are different from each other. In this study, the conceptual designs of 8 DMS are modeled with UML [15] class diagrams. However, in this paper, two representatives are selected just to illustrate the analysis.

### 2.1 M-Files

M-Files is one of the best document management systems, proposed by the Internet reviews. M-Files has two important parts. One part is M-Files Server and the other is M-Files Client, which is client-server architecture.

The M-files meta model can be seen in Figure 2.1. The M-Files has a "vault" concept which has different types of documents and users. A vault is a centralized storage location for documents and other objects. The vault has metadata types, workflows, role/groups and users as concepts. A user sees the document vault as a directory on their local computer's M-Files drive.

One important feature of M-Files is its metadata concept [6]. The Metadata is described as data about data. It means that descriptive data related to the context can be defined. Metadata concept has four important entities, which are Employees, Customers, Projects and Documents. All saved objects in M-Files are described with metadata. In traditional document management systems, metadata cannot be defined according to the needs of the organization. Document vault, object type, class groups, classes and properties are defined as metadata components. The explanations of these terms are defined shortly below [7].

The Object type is the equivalent concept for metadata. Document, Employee, Customer and Project are object types. All objects are saved as object type in M-Files. Object types other than document or document collection can be considered as data registries within the system. Class group is used for gathering similar classes together. Sales, Meetings, Production etc. can be given as examples. Classes are used to classify object types into smaller entities. Classes inform what the type of this document is. Agreement, Report, Memo and Agenda etc. can also be given as examples. Properties are defined class-specific.

The workflow concept is supplies a functionality for M-Files users. The M-Files workflow feature enables modeling object lifecycles according to real world processes. The workflow is grouped into states that correspond to the working stages of the documents or other objects. The M-Files administrator can easily define workflows to meet company requirements. A workflow user can create workflow in M-Files. A workflow user has a worklist. This worklist has many workitems. The cases enables to enact specific objectives of workflow. Tasks have states in workflow. The general workflow structure is defined with these entities for some users in M-Files meta model.

The other important part in vault is Roles/Groups, which have one or more access control lists (ACL). M-Files have defined users which have user groups. These user groups can create by any user. User groups have internal users and external users. Users and user groups have mail groups, which are used for sharing all documents between other users. User and user groups have various roles on system management, including management rights and permissions for M-Files users. At the same time, users use and manage the vault.

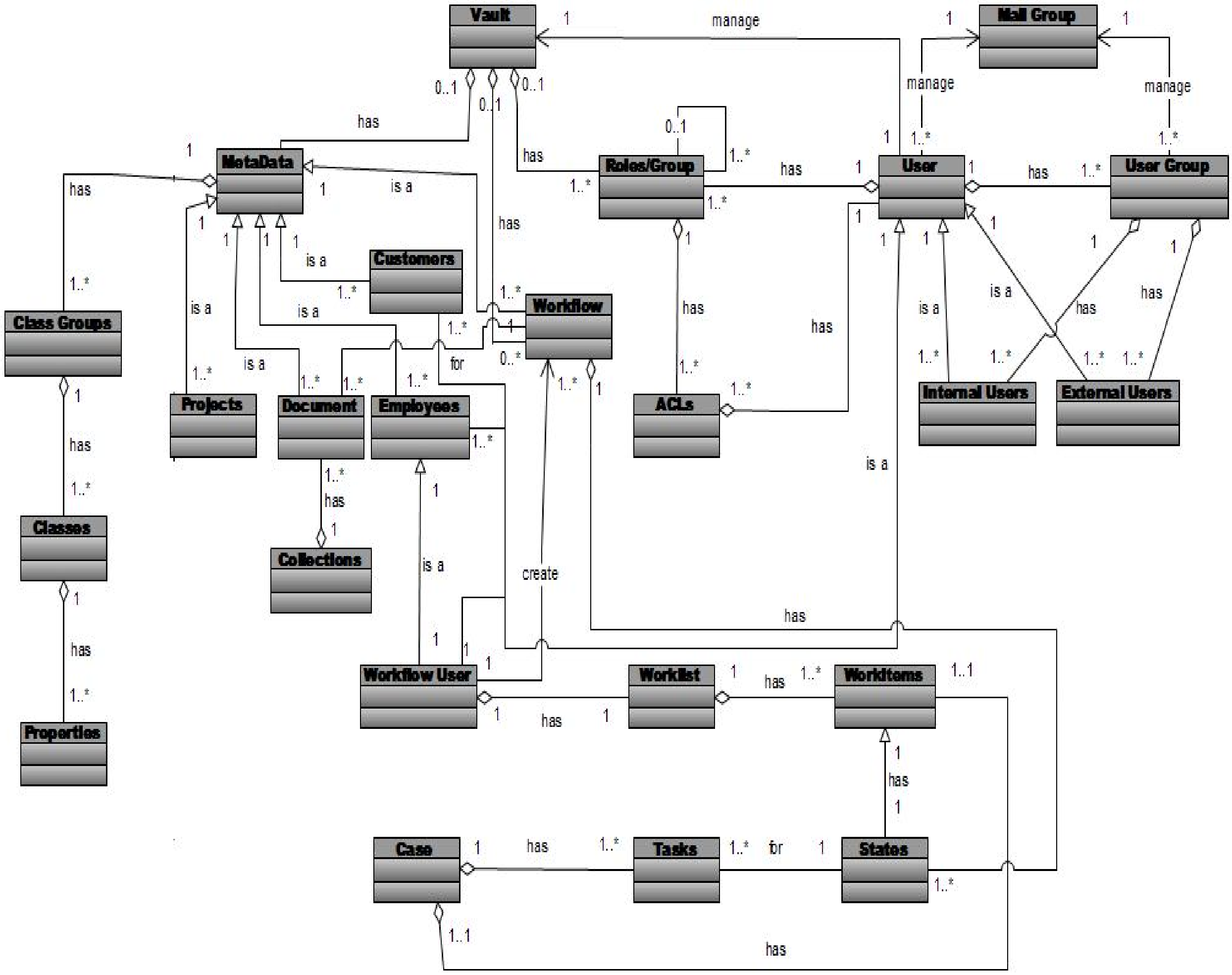

Figure 2.1. The M-Files Meta Model

## 2.2 Paperport

Paperport is document management software that helps to organize access, share and manage documents and image files on personal computers [11]. It is an individual desktop DMS, but it has a high usage point at Internet reviews. Because of this, the Paperport meta model is also examined. Paperport has a simple meta model unlike others. It has Roles, User and Mail Groups fundamentally. Documents are defined as items in meta model. Item groups may have items as tree view structure. All documents and folders are uploaded directly. It stores all documents as a tree view hierarchical structure. A file cabinet or vault is not defined such systems do not exist in meta model. Moreover, it has not workflow feature. There are only folder and document objects. It has a database for storage but this database is not a vault or a file cabinet. There is the user who can manage the documents and folders. A user may have user groups which has some roles, however, admin, author or other discriminations do not exist. Everyone is defined as user. Furthermore, mail groups also provide document sharing with other people like other systems. You can examine this Meta model on Figure 2.2. [16].

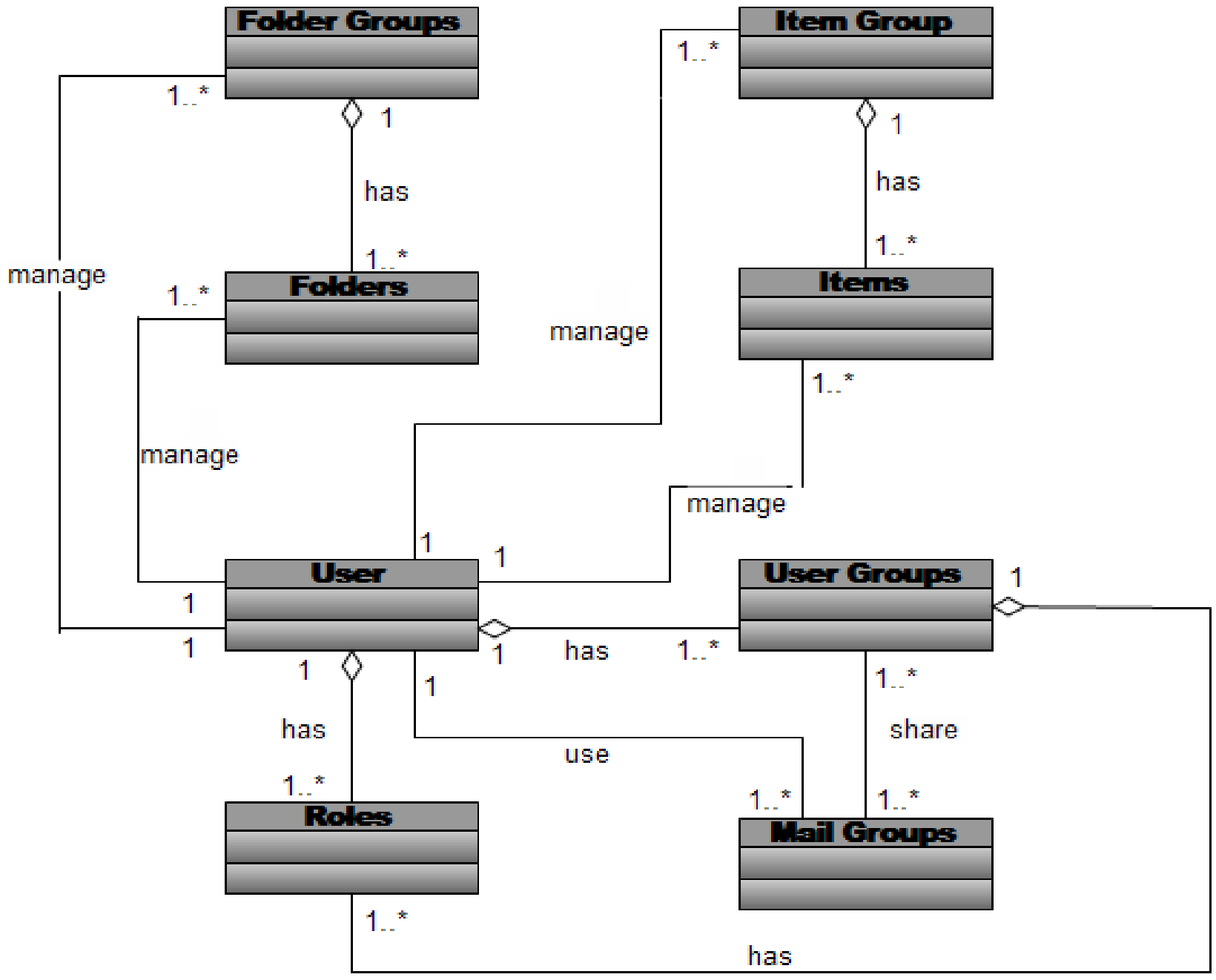

Figure 2.2. The Paperport Meta Model.

M-Files and Paperport are different from each other from many aspects. Two main groups of DMSs are seen based on similar meta models. In this paper, two representatives for meta model analysis are selected. Blue Doc Point, Speedy Organizer, Iso Tracker, Doc Point, Paperless and File Center Pro meta model analysis are found in [16]. However, the overall conclusion of comparison of these systems is provided in this paper.

The most important criteria while comparing those meta models are the number of concepts, relationship types, cardinalities and attributes. From the eight meta models, it is observed that M-Files, Blue-Doc Point, Speedy Organizer and Doc-Point have more concepts than File Center Pro, Iso Tracker, Paperless and Paperport. This means that the first mentioned group supports a wider scope. If variety of concepts is available, a more adaptable system can be defined. Therefore, it can be said that M-Files is an adaptable system according to variety of concepts. Blue-Doc Point, Speedy Organizer and Doc-Point scope of concepts come after M-Files. Therefore, these systems are more flexible and adaptable than other DMSs.

On the other hand, some of DMSs have vault or file cabinet concepts. This is the flexibility factor for them. M-Files, Speedy Organizer, Doc-Point, Iso Tracker, File Center and Paperless have this concept as the root storage meta data concept. There is user group concept at some models. This concept is important for flexibility and adaptability. Because, normal users can manage other user groups via this concept.

Hierarchical tree view structure provides more flexibility. This supported by M-Files, Speedy Organizer, Iso Tracker and Blue-Doc Point. These also have workflow features and user groups, mail groups and metadata objects.

The workflow feature is very important from the point of view of flexibility and adaptability factor for comparison of document management systems. Paperport, Paperless, File Center Pro and Doc-Point does not have workflow feature. These models only have tree view metadata structure instead of workflows. Therefore, these models can be less flexible and adaptable than others according to these concepts.

Second important criteria for comparison of meta models are the number of relationships which determines the structural density. The existence of different relationship types between two concepts provides more flexibility. For example, M-Files metadata concept has both inheritance and association relationship with other related entities. All of concepts in M-Files have different relationships with other concepts. Speedy Organizer, Blue-Doc Point and Doc-Point have similar relationships (both inheritance and association relationships). Other models have only association relationships and directed associations with other concepts. Therefore, they can be thought as less flexible than M-Files, Speedy Organizer and Doc-Point according to relationship types.

M-Files appeared to be the most supportive document management system. This result is obtained according to concepts, relationships, and cardinalities and attributes comparisons. It can be said that these eight models divided into two main groups. The first group has metadata hierarchical structure, User-User-Groups, Roles / Groups and Workflow Feature. This group includes M-Files, Blue-Doc Point, Speedy Organizer and Iso Tracker. The second group does not have hierarchical metadata structure. They have tree view structure for storing documents. They do not have workflow feature. They have less relationships and concepts than others. These models are File Center Pro, Paperless, Paperport and Doc-Point. For concept comparisons, table 1 can be viewed.

To obtain the comparison table, 8 DMS products have been installed, configured, run and they are examined in detail by using their user interface and documentation. Identified concepts and relations of DMS's have been modeled. Obtained meta models showed various underlying conceptual designs. Then the models are examined, different and overlapping concepts are identified and consolidated in the table. If a DMS has a concept, this has been shown with plus (+) sign. If it has not any related concept, this has been seen with minus (-) sign. Plus and minus signs indicate whether the concept is exists those models or not. This analyisis included 8 DMS meta models in [16].

Table 1. Comparison of concepts for DMSs

| **Entity Types** | **MFiles** | **BlueDoc** | **Speedy Organizer** | **DocPoint** | **FileCenter Professinal** | **Paperless** | **Paperport** | **ISOTracker** |
|---|---|---|---|---|---|---|---|---|
| | + | - | + | + | + | + | - | + |
| Advanced/Standard/Contacts Cabinet | - | - | - | - | + | - | - | - |
| Roles / Groups | + | + | - | - | + | + | + | + |
| Vault/FileCabinet/Libraries/Modules | + | + | + | + | - | - | - | + |
| Mail Groups / Mail Cabinet | + | + | + | - | + | + | + | - |
| User | + | + | + | + | + | + | + | + |
| Manager / Admin / Auther | - | - | + | - | + | - | - | + |
| User Groups | + | + | - | - | - | - | + | + |
| Internal / External Users | + | - | - | - | - | - | - | - |
| Workflow / Alerts | + | + | + | - | - | - | - | + |
| Workflow Role | - | + | - | - | - | - | - | - |
| Workflow Users / Performer | + | + | - | - | - | - | - | - |
| Worklists | + | - | - | - | - | - | - | - |
| Workitems | + | - | + | - | - | - | - | - |
| Cases / Tasks / States | + | + | - | - | - | - | - | - |
| Activities / Processes | - | + | - | - | - | - | - | + |
| Metadata | + | - | - | - | - | - | - | - |
| Class Group | + | - | - | - | - | - | - | - |
| Classes | + | - | - | - | - | - | - | - |
| Properties | + | - | - | + | - | - | - | - |
| Employees | + | - | - | - | - | - | - | + |
| Customers / Suppliers | + | - | - | - | - | - | - | + |
| Documents | + | + | + | + | - | - | - | + |
| Incoming /Internal/Outgoing Docs. | - | - | - | + | - | - | - | - |
| Collections | + | - | - | - | - | + | - | - |
| Projects | + | - | - | - | - | - | - | - |
| Folders / Folder Groups | - | + | + | + | + | + | + | + |
| Subfolders | - | - | - | - | + | - | - | - |
| Files | - | - | + | - | + | - | - | + |
| Templates / Header | - | - | - | + | + | - | - | - |
| Working Groups | - | - | - | - | - | - | - | - |
| Drawers / Droplets | - | - | - | - | + | + | - | - |
| Records | - | - | + | - | - | - | - | - |
| Inactive / Active records | - | - | + | - | - | - | - | - |
| Item / Item Groups | - | - | - | - | - | + | + | - |
| Modules | - | - | - | - | - | - | - | + |
| Complaints | - | - | - | - | - | - | - | + |
| Competency | - | - | - | - | - | - | - | + |
| Connectors | - | - | - | + | - | - | - | - |
| Communication | - | - | - | + | - | - | - | - |
| Contact / Contact Groups | - | - | - | + | - | - | - | - |
| Inbox | - | - | - | - | + | - | - | - |

## 3. FLEXIBILITY AND HUMAN ORIENTATION OF DOCUMENT MANAGEMENT SYSTEMS BASED ON TUNING MEASURES

Flexibility is defined as “The ability of the concept to proactively, reactively or inherently embrace change in the timely manner through its components and its relationships with environment”. [3]

The tuning measures are generated through two important theoretical backgrounds. This theoretical framework is composed of Job Characteristic Model and Assignment and Synchronization Policies [1]. There are 21 important tuning measures according to Vanderfeesten. These measures ensure that the system is becoming more human-oriented and user-friendly. In general, the measures provide more autonomy or more self-determination system for document management system’s user.

Document management systems according to these tuning measures are examined to determine whether they are worker-oriented or not. “Which document management system is more worker-oriented?”, “Which one is user friendly?”, “Which one has more work autonomy?” the answers to these questions will be studied.

The comparison of document management products are based on these 21 tuning measures. These tuning measures are explained in details in Vanderfeesten article [4]. The comparison of these systems based on tuning measures is shown following table. For 8 DMS products, all tuning measures have been investigated through both product itself, user documentation and meta models. The plus sign (+) shows that whether the product support this measure or not.(+/-) sign shows also that this product provides pratial support of this measure. Minus sign (-) shows that this product does not have any support for measures. Since tuning measures are more associated with product than their models, all DMS products have been examined in detail in order to determine all signs at table 2. But, meta models have also been used in investigation for tuning measures criteria. All measures have been studied through both products and models. The results which are related to the flexibility and human-orientation are shown at table 2. On the other hand, all results at table 2 will be explained through the M-Files example. All results in table 2 have been obtained like M-Files examination example for tuning measures. All results can be seen from table 2.

Table 2. Comparison of tuning measures for DMSs

| **Tuning Measures** | M-Files | BlueDoc | Speedy Organizer | Doc Point | File Center Pro | Paperless | Paperport | ISO Tracker |
|---|---|---|---|---|---|---|---|---|
| **APPEAR** | + | + | + | +/- | +/- | +/- | +/- | + |
| **BATCH** | + | - | + | + | - | + | + | + |
| **CASEMAN** | + | + | + | - | - | - | +/- | + |
| **HISTORY** | - | - | - | - | - | - | - | - |
| **ITEMS** | + | - | + | - | - | - | + | + |
| **PREFS** | + | + | - | + | + | - | - | - |
| **PRIVPULL** | + | + | + | + | + | + | + | + |
| **RANDOM** | - | + | - | - | - | + | + | +/- |
| **RANKING** | - | - | + | - | - | - | - | + |
| **REDIRECT** | + | + | + | + | + | + | + | + |
| **REJECT** | + | + | + | - | - | - | - | + |
| **RELEASE** | + | + | + | + | + | + | + | + |
| **REPLAN** | +/- | +/- | +/- | - | - | - | - | +/- |
| **RESUBMIT** | + | + | + | + | - | - | - | + |
| **RESULT** | + | + | + | - | - | - | + | + |
| **ROUTE** | + | + | + | + | + | - | - | - |
| **SHPULL** | - | + | + | - | +/- | - | - | - |
| **STATMOD** | - | +/- | - | - | - | - | - | - |
| **TARGET** | + | + | + | +/- | - | - | - | +/- |
| **TEAMBAT** | + | + | - | - | + | - | + | + |
| **TEAMWI** | + | + | - | - | + | - | +/- | + |

M-Files meta model structure has been explained above. Now, their flexibility will be examined shortly. Tuning measures will be used while investigating flexibility and human-orientation. These measures are shown at table-2 above. All measures through M-Files will be studied as example.

M-files support [APPEAR] measure. Any employee can change workitem's appearance according to their own preference. In worklist, the workitems are adjusted. M-Files store all document types in a vault. Any document can be chosen from this vault. Because of this, employee can select the execution order of work items within this batch. It provides [BATCH] measure. It is case oriented system. Therefore, [CASEMAN] measure is supported. [HISTORY] and [PREFS] measures are not supported. A new case offer through old cases is not supported. A worklist may have one or more workitems. [ITEMS]

An employee decides the work items which will work on and he/she can select workitem from private worklist. Workitem's order is not random. Thus, it has not [RANDOM] property. It has not any employee ranking degree. It has not [RANKING]. There is an object which is called "Mail Group". It provides to share all documents between all users. [REDIRECT] means sharing data. So, it has this measure.

All users have some authorization rights on the main library. One of the rights is to destroy any objects. So, [REJECT] is provided by this system. A new work item can release in worklist. [RELEASE]. You do not "re-plan" on workitems. When a work item has any changes, employee must receive feedback about this workitem. This property is called [RE-SUMMIT] and it's supported. Employees can see workflow's result. [RESULT]. There is a "ACLs" object in meta model. This object provides to user some private rights. These rights refer to [ROUTE] measure in tuning measures and M-Files has it?

As seen from meta model, there is a one-to-one relationship between workflow user and worklist. Therefore, shared worklist cannot be created. It is not [SH PULL]. It has no design diagram about workflows. There is no [STAT MOD] property. In the metamodel, there is a "User Group" concept. One can create mutual sharing group. Worklist can be divided according to user group. So, [TEAM BATCHES] is supported in M-Files. Users who have different authorizations can run an activity together. Therefore, M-Files has [TEAM WI] measure.

From this evaluation, M-Files can be stated as a flexible and human-oriented document management system. It has a large scope and it has most of important tuning measures. While examining other systems, similar research techniques are used. These tuning measures are controlled step by step with both their meta models and software. These results are shown at above table-2. According to this table, M-Files, Speedy Organizer, Blue-Doc Point and Iso Tracker have most of the tuning measures. Therefore, these document management systems are more human oriented and user friendly than others. Moreover, Paperless has the least number of measures than other systems.

## 4. CONCLUSION

Eight document management products for research in this paper have high usability degree in the market. Their meta models are examined according to concepts, relationships. Four of the systems have workflow feature different from the other systems. This feature is one of the flexibility factors. According to meta model comparisons, M-Files is the most supportive document management system between the other systems. These eight models can be divided into two main groups. The first group has metadata hierarchical structure and the second group does not have hierarchical metadata structure. They have tree view structure in order to store documents. They do not have workflow feature. They have less relationships and concepts than others. This comparison is shown at table 1 above.

The other evaluation for these systems is that comparing flexibility and human-orientation factors. There are 21 important tuning measures for evaluation of flexibility and human-orientation. Tuning measures are used step by step both meta models and systems. According to these factors from table-2, M-Files, Blue Doc, Speedy Organizer and Iso Tracker are more flexible and human-oriented systems than the others. Human-orientation and flexibility concepts are related to positive tuning measures numbers. If many tuning measures are positive, the more flexible and human-oriented the system is.

# APPENDIX

## List of Tuning Measures which is proposed by Vandeerfeesten. (Vandeerfeesten, 2006)

**[APPEAR]**Give employees the opportunity to adjust the appearance of work items in their worklists to their own preferences: FCFS, earliest due date, random, etc. (Here we assume the assignment of work items is in a pull manner and the worklist is private).

**[BATCH]** Offer an employee ‘‘batches’’ of work items. In this way the batch is pushed, but the employee can choose the order of execution of work items within this batch. (Here we assume the worklist is private).

**[CASEMAN]** Case management: let an employee work on the same case as much as possible.

**[HISTORY]** Offer a variety in work items to an employee. Remember the kind of work items an employee has executed and decide, based on this history, what kind of new work items will be offered to him or her.

**[ITEMS]** If possible, show more than one work item on the worklist of an employee, even if a push mechanism is used.

**[PREFS]** Keep up with the kind of activities an employee likes and make sure he or she will get more of this kind of activities (and less of activities (s)he does not like).

**[PRIV PULL]** Let an employee choose work items from the private worklist himself/herself: pull-mechanism.

**[RANDOM]** The queuing of work item sin the worklist should be random.

**[RANKING]** Make available an employee’s place within the ranking of good employees (for instance ‘‘hard working’’, ‘‘producing high quality work’’).

**[REDIRECT]** Give employees the possibility to send a work item to another employee, who is better in performing the job, who has more knowledge about the case who is not busy, etc.

**[REJECT]** Give employees the possibility to reject a work item (with a valid reason) and return it to the workflow enactment service.

**[RELEASE]** Release a new work item directly. (Time of notification is upon availability).

**[REPLAN]** Do not ‘‘re-plan’’ work items by workflow enactment service.

**[RESUBMIT]** When a work item has to be performed again after a (negative result of a) check, return it to the same employee to execute it again.

**[RESULT]** Give each employee authorization to view the final decision or result of a case in the process.

**[ROUTE]** Give an employee the possibility to check the progress and route of a case during the process (dynamic aspect).

**[SH PULL]** Use a shared worklist, from which an employee can choose him/her:

Pull-manner.

**[STAT MOD]** Design a possibility for the employee to examine the static process model in a comprehensible way (static aspect).

**[TARGET]** Show an employee if he or she works hard enough, if he or she is satisfying the targets.

**[TEAM BAT]** Create ‘‘team batches’’ of work items. A team of employees (having the same competences/role) can divide the work according to their own preferences. (Here we assume the

allocation mechanism is manual, but is not necessarily controlled by a team leader or manager). This idea is quite similar to the concept of a self-managing team, which is one of the hot items in organizational psychology.

**[TEAM WI]** Create ‘‘team work items’’. Employees (with different competences) have to cooperate to execute an activity.

**Authors**

Ö. Özgür Tanrıöver

Ö. Özgür Tanrıöver received Science and Technology Policy MSc (2001), Information Systems MSc (2002) and PhD (2008) degrees from Informatics Institute, Middle East Technical University (METU), and Ankara, Turkey. He was a research assistant at the Center for Science and Technology Policy Research - METU between 1998 and 2005. Between 2005 and 2011, he has been a certified (CISA) Information Systems Auditor in the Information Management Department at the Banking Regulation Agency of Turkey. Since 2012, he is a professor in Computer Engineering Department of Ankara University. His current research interests are Software Quality, Information Technology Management and Conceptual Modeling.

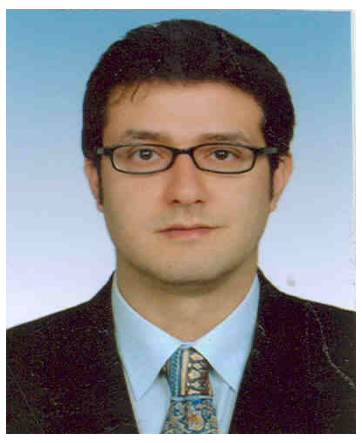

Harika Asılı

Harika Asılı is an undergraduate senior student at Ankara University, Ankara, Turkey. She has been studying Computer Engineering Department since 2010. Meanwhile, she has been studying Economics, in the Department of Political Sciences Faculty at Ankara University since 2011 as double major undergraduate program student. She worked at Schneider Electric Inc. , Vestel Electronic Inc. and Denizbank Inc. in summer 2012 and 2013 as an intern engineer. Since February 2014, she has been working at Technology Security Operations Department as a part-time computer engineer in Vodafone Telecommunication Inc. at Ankara. She is studying on graduation thesis about conceptual software modelling on document management systems and their flexibility and human-orientation factors.

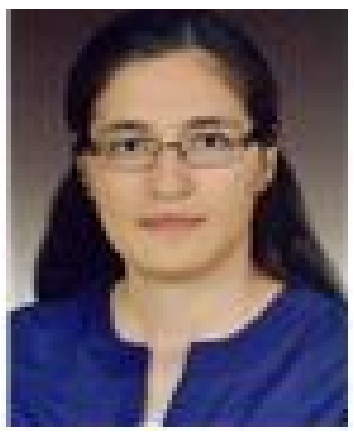